\documentclass[twocolumn,showpacs,aps,prl]{revtex4}
\newcommand{\half}{\frac{1}{2}}

\newcommand{\frth}{\frac{1}{4}}
\newcommand{\sxth}{\frac{1}{6}}
\begin{document}
\title{
The Higgs scalar field with no massive Higgs particle}
\author{R. K. Nesbet }
\affiliation{
IBM Almaden Research Center,
650 Harry Road,
San Jose, CA 95120-6099, USA }
\date{\today}
%\begin{center}For {\em PhysRevLett} \end{center}
\begin{abstract}
%***************** VM screen width ************************************%
The postulate that all massless elementary fields have conformal Weyl
local scaling symmetry has remarkable consequences for both cosmology 
and elementary particle physics.  Conformal symmetry couples scalar and 
gravitational fields. Implications for the scalar field of a conformal 
Higgs model are considered here.  The energy-momentum tensor of a 
conformal Higgs scalar field determines a cosmological constant. It has 
recently been shown that this accounts for the observed magnitude 
of dark energy.  The gravitational field equation forces the energy
density to be finite, which precludes spontaneous destabilization of 
the vacuum state.  Scalar field fluctuations would define a Higgs 
tachyon rather than a massive particle, consistent with the ongoing 
failure to observe such a particle. 
\end{abstract}
%\pacs{11.15.-q}{Gauge field theories}
%\pacs{14.80.Bn}{Standard-model Higgs bosons}
%\pacs{04.20.Cv}{General relativity, fundamental problems}
%\pacs{98.80.-k}{Cosmology}
\maketitle
\section{Introduction}
The standard model of spinor and gauge boson fields has higher symmetry
than does Einstein gravitational theory\cite{MAN06}. 
For massless fields with definite conformal character 
action integrals are invariant under local Weyl (conformal) scaling,
$g_{\mu\nu}(x) \to g_{\mu\nu}(x) e^{2\alpha(x)}$\cite{MAN06}.
A conformal energy-momentum tensor is traceless, while the Einstein 
tensor is not.  
\par Compatibility can be imposed in gravitational theory by replacing 
the Einstein-Hilbert field action by a uniquely determined action 
integral $I_g$ constructed using the conformal Weyl tensor\cite{MAN06}. 
Conformal gravity accounts for anomalous galactic rotation velocities 
without invoking dark matter\cite{MAN06}.  Relativistic phenomenology 
at the distance scale of the solar system is preserved.  
\par An inherent conflict between gravitational and elementary particle 
theory is removed if all massless elementary fields have conformal 
symmetry.  Standard cosmology\cite{DOD03} postulates uniform, isotropic 
geometry, described by the Robertson-Walker (RW) metric tensor.
In RW geometry, conformal gravitational 
${\cal L}_g$ vanishes identically\cite{MAN06}, but the residual 
gravitational effect of a conformal scalar field is consistent with 
Hubble expansion\cite{MAN06}, dominated in the current epoch by dark 
energy, with negligible spatial curvature\cite{NES11,KOM09}. 
\par
In electroweak theory, the Higgs mechanism introduces an SU(2) doublet 
scalar field $\Phi$ that generates gauge boson mass\cite{PAS95,CAG98}.
Postulating universal conformal symmetry for massless elementary 
fields, these two scalar fields can be identified\cite{NES10}.
Lagrangian density ${\cal L}_\Phi$ for conformal scalar field 
$\Phi(x)\to\Phi(x)e^{-\alpha(x)}$ includes a term dependent on Ricci 
scalar $R=g_{\mu\nu}R^{\mu\nu}$, where $R^{\mu\nu}$ is the 
gravitational Ricci tensor\cite{MAN06}. 
In uniform, isotropic geometry this determines a modified 
Friedmann cosmic evolution equation\cite{NES11} consistent with
cosmological data back to the microwave background epoch\cite{KOM09}.
\par Implications for the standard electroweak model are examined here. 
The Higgs model Lagrangian density contains
$\Delta{\cal L}_\Phi=(w^2-\lambda\Phi^\dag\Phi)\Phi^\dag\Phi$, where
$w^2$ and $\lambda$ are undetermined positive constants\cite{CAG98}.  
Units here set $\hbar=c=1$.
Lagrangian term $\lambda(\Phi^\dag\Phi)^2$ is conformally covariant. 
$w^2\Phi^\dag\Phi$ breaks conformal symmetry, but can be generated 
dynamically\cite{NES10}.  Conformal symmetry requires a term 
$-\sxth R\Phi^\dag\Phi$\cite{MAN06}.  Empirical cosmological 
$R>0$\cite{NES11}, so $-\sxth R$ and $w^2$ have opposite signs. A
consistent theory must include $(w^2-\sxth R)\Phi^\dag\Phi$\cite{NES11}.
\par The conformal scalar field equation has exact solutions such that
$\Phi^\dag\Phi=\phi_0^2=(w^2-\sxth R)/2\lambda$, if this ratio is
positive and $R$ is treated as a constant.
Only the magnitude of $\Phi$ is determined. For $\phi_0^2>0$,
a modified Friedmann cosmic evolution equation has been 
derived\cite{NES11} and solved to determine cosmological parameters. 
The residual constant term in conformal 
energy-momentum tensor $\Theta^{\mu\nu}_\Phi$ defines a
cosmological constant (dark energy)\cite{MAN06,NES11}.
Nonzero $\phi_0^2$ produces gauge boson masses\cite{CAG98}. 
\par Conformal theory identifies $w^2$ with the empirically positive  
cosmological constant\cite{NES11}, but does not specify the algebraic
sign of parameter $\lambda$.  For the Higgs mechanism, condition  
$\phi_0^2=(w^2-\sxth R)/2\lambda>0$ requires the sign of $\lambda$ to
agree with $w^2-\sxth R$.  The scalar field energy density determined 
by the coupled equations derived here is necessarily finite for any 
real value of $\lambda$. This precludes destabilization of the vacuum.  
\par Fluctuations $\delta\phi\to0$ about an exact solution of the scalar
field equation satisfy
$\partial_\mu\partial^\mu\delta\phi\to-4\lambda\phi_0^2\delta\phi$.
If $\lambda>0$ this is a Klein-Gordon equation with
$m_H^2=4\lambda\phi_0^2=2(w^2-\sxth R)$, 
which defines a Higgs boson\cite{CAG98} if $R<6w^2$.
In the conformal Higgs model, empirical values of parameters $w^2$, 
$R$, and $\phi_0^2$ determine parameter $\lambda$.  It is 
argued here that these parameters, now well-established from 
cosmological and electroweak data, imply $\lambda<0$, 
consistent with an earlier formal argument\cite{MAN06}. 
Hence fluctuations of a conformal Higgs scalar field do not satisfy a
Klein-Gordon equation.  This rules out a standard Higgs particle of any
real mass.  Negative $m_H^2$, or finite pure imaginary mass, would
define a tachyon\cite{FEI67}, if such a particle or field could exist, 
and might justify an experimental search for such a tachyon.
%***************** VM screen width ************************************%

\section{The modified Friedmann equation}
\par In cosmological theory, a uniform, isotropic universe is described
by Robertson-Walker (RW) metric\\
$ds^2=dt^2-a^2(t)(\frac{dr^2}{1-kr^2}+r^2d\omega^2)$,
if $c=\hbar=1$
and $d\omega^2=d\theta^2+\sin^2\theta d\phi^2$.
Gravitational field equations are determined by Ricci tensor
$R^{\mu\nu}$ and scalar $R$.  
The RW metric defines two independent functions
$\xi_0(t)=\frac{\ddot a}{a}$ and
$\xi_1(t)=\frac{{\dot a}^2}{a^2}+\frac{k}{a^2}$,
such that $R^{00}=3\xi_0$ and $R=6(\xi_0+\xi_1)$.
The field equations reduce to Friedmann equations for 
scale factor $a(t)$ and Hubble function $h(t)=\frac{{\dot a}}{a}(t)$.
\par If the scalar field required by Higgs symmetry-breaking has 
conformal symmetry, its action integral $I_\Phi$ must depend on the 
Ricci scalar, implying a gravitational effect.  
Because conformal gravitational action integral $I_g$ vanishes 
identically in RW geometry\cite{MAN06}, it is consistent to assume that 
uniform cosmological gravity is determined by this scalar field. 
\par Including term $(w^2-\sxth R)\Phi^\dag\Phi$ in 
${\cal L}_\Phi$\cite{NES11},
the field equation for scalar $\Phi$ is
$\partial_\mu\partial^\mu\Phi=
 (w^2-\sxth R-2\lambda\Phi^\dag\Phi)\Phi$. \\
Generalizing the Higgs construction, and neglecting the cosmological
time derivative of $R$, constant $\Phi=\phi_0$ is a global solution if
$\phi_0^2=\frac{1}{2\lambda}(w^2-\sxth R)$.  Evaluated for 
this field solution, 
${\cal L}_\Phi=\phi_0^2(w^2-\sxth R-\lambda\phi_0^2)
=\half\phi_0^2(w^2-\sxth R)$. 
\par Variational formalism of classical field theory\cite{NES03} 
is easily extended to the context of general relativity\cite{MAN06}. 
The metric functional derivative
$\frac{1}{\sqrt{-g}}\frac{\delta I}{\delta g_{\mu\nu}}$ 
of generic action integral $I=\int d^4x\sqrt{-g}{\cal L}$
is $X^{\mu\nu}=x^{\mu\nu}+\half g^{\mu\nu}{\cal L}$, 
if $\delta{\cal L}=x^{\mu\nu}\delta g_{\mu\nu}$.
The energy-momentum tensor is $\Theta^{\mu\nu}=-2X^{\mu\nu}$.
Varying $g_{\mu\nu}$ for fixed scalar field solution $\Phi$, 
metric functional derivative
\begin{eqnarray}
 X_\Phi^{\mu\nu}=
\sxth R^{\mu\nu}\Phi^\dag\Phi+\half g^{\mu\nu}{\cal L}_\Phi
\nonumber \\
=\sxth\phi_0^2(R^{\mu\nu}-\frth Rg^{\mu\nu}+\frac{3}{2}w^2g^{\mu\nu}) 
\end{eqnarray}
implies modified Einstein and Friedmann equations\cite{NES11}.
\par The gravitational field equation driven by energy-momentum tensor
$\Theta_m^{\mu\nu}=-2X_m^{\mu\nu}$ for uniform matter and radiation is
$X_\Phi^{\mu\nu}=\half\Theta_m^{\mu\nu}$.
Since $\Theta_m^{\mu\nu}$ is finite, determined by fields independent of
$\Phi$, $X_\Phi^{\mu\nu}$ must be finite, regardless of any parameters 
of the theory.  This precludes spontaneous destabilization of the 
conformal Higgs model.
\par Defining ${\bar\kappa}=-3/\phi_0^2$ and
${\bar\Lambda}=\frac{3}{2}w^2$, the modified Einstein equation is 
\begin{eqnarray}
R^{\mu\nu}-\frth Rg^{\mu\nu}+{\bar\Lambda}g^{\mu\nu}
 =-{\bar\kappa}\Theta_m^{\mu\nu}.
\end{eqnarray}
Traceless conformal tensor $R^{\mu\nu}-\frth Rg^{\mu\nu}$ here replaces
the Einstein tensor of standard theory\cite{NES11}. 
Cosmological constant ${\bar\Lambda}$ is determined by Higgs parameter
$w^2$.  Nonstandard parameter ${\bar\kappa}<0$ is
determined by the scalar field\cite{MAN06,NES11}.
For energy density $\rho=\Theta_m^{00}$ this implies
$-\frac{2}{3}(R^{00}-\frth R)= \xi_1(t)-\xi_0(t) 
    =\frac{2}{3}({\bar\kappa}\rho+{\bar\Lambda})$.
Hence uniform, isotropic matter and radiation determine the 
modified Friedmann cosmic evolution equation\cite{NES11} 
\begin{eqnarray}
\xi_1(t)-\xi_0(t)=
\frac{{\dot a}^2}{a^2}+\frac{k}{a^2}-\frac{\ddot a}{a}=
    \frac{2}{3}({\bar\kappa}\rho+{\bar\Lambda}). 
\end{eqnarray}
\par Because the trace of $R^{\mu\nu}-\frth Rg^{\mu\nu}$ is identically
zero, a consistent theory must satisfy the trace condition
$g_{\mu\nu}{\bar\Lambda} g^{\mu\nu}=
 4{\bar\Lambda}=-{\bar\kappa}g_{\mu\nu}\Theta_m^{\mu\nu}$.
From the definition of an energy-momentum tensor, this is just
the trace condition satisfied in conformal theory\cite{MAN09},
$g_{\mu\nu}(X_\Phi^{\mu\nu}+X_m^{\mu\nu})=0$.  Vanishing trace 
eliminates the second Friedmann equation derived in standard theory.
Although the $w^2$ term in $\Delta{\cal L}_\Phi$ breaks conformal
symmetry, a detailed argument shows that the trace
condition is preserved\cite{NES10}.

\section{Fits to cosmological data}
\par The modified Friedmann equation determines dimensionless
scale parameter $a(t)=1/(1+z(t))$, for redshift $z(t)$, and
function $h(t)=\frac{{\dot a}}{a}(t)$ in units of current
Hubble constant $H_0=$70.5 km/s/Mpc\cite{KOM09}, such that 
$z=0, a=1, h=1$ at present time $t_0$.
Distances here are in Hubble units $c/H_0$.
\par The modified Friedmann equation depends on nominally constant
parameters, fitted to cosmological data for $z\leq z_*$: 
$\alpha=\frac{2}{3}{\bar\Lambda}=w^2>0$,
$k\simeq0$, $\beta=-\frac{2}{3}{\bar\kappa}\rho_m a^3>0$, and
$\gamma=3\beta/4R_b(t_0)$.
$z_*=1090$ here characterizes the cosmic microwave background, at $t_*$,
when radiation became decoupled from matter.
$\frac{4}{3}R_b(t)$ is the 
ratio of baryon to radiation energy densities.  
Empirical value $R_b(t_0)=688.6$\cite{KOM09} is assumed.
Scaled energy densities $\rho_m a^3$ and $\rho_r a^4$, for matter
and radiation respectively, are constant. 
In the absence of dark matter, $\rho_m\simeq\rho_b$, the baryon density.
\par The parametrized modified Friedmann equation is
\begin{eqnarray}
\frac{{\dot a}^2}{a^2}-\frac{{\ddot a}}{a}=
 -\frac{d}{dt}\frac{{\dot a}}{a}={\hat\alpha}=
 \alpha-\frac{k}{a^2}-\frac{\beta}{a^3}-\frac{\gamma}{a^4}.
\end{eqnarray}
Dividing this equation by $h^2(t)$ implies dimensionless sum rule 
\begin{eqnarray}
\Omega_m(t)+\Omega_r(t)+\Omega_\Lambda(t)+\Omega_k(t)+\Omega_q(t)=1,
\end{eqnarray}
where 
$\Omega_m(t)= \frac{2}{3}\frac{{\bar\kappa}\rho_m(t)}{h^2(t)}<0$,
$\Omega_r(t)= \frac{2}{3}\frac{{\bar\kappa}\rho_r(t)}{h^2(t)}<0$,
$\Omega_\Lambda(t)=\frac{w^2}{h^2(t)}>0$,
$\Omega_k(t)=-\frac{k}{a^2(t)h^2(t)}$, and
$\Omega_q(t)=\frac{{\ddot a}a}{{\dot a}^2}=-q(t)$.
In contrast to the standard sum rule, $\Omega_m$ and $\Omega_r$ are 
negative, while acceleration parameter $\Omega_q(t)$ appears explicitly.
\par Hubble expansion is characterized for type Ia supernovae by scaled
luminosity distance $d_L$ as a function of redshift $z$.
Here $d_L(z)=(1+z)d_z$, for geodesic distance $d_z$ corresponding to
$r_z =\int cdt/a(t)$, integrated from $t_z$ to $t_0$.  In curved space 
(for $k<0$), $d_z =\frac{\sinh(\sqrt{-k}r_z)}{\sqrt{-k}}$.
In the standard $\Lambda CDM$ model\cite{DOD03}, radiation density and
curvature $\Omega_k$ can be neglected in the current epoch ($z\leq 1$).
This reduces the sum rule to $\Omega_\Lambda+\Omega_m=1$.  Empirical
value $\Omega_\Lambda=0.726$ forces $\Omega_m$ to be much larger than 
can be accounted for by observed matter, providing a strong argument for
dark matter.  Mannheim\cite{MAN03,MAN06} questioned this implication, 
and showed that observed luminosities could be fitted equally well 
for $z\leq 1$ with $\Omega_m=0$, using the standard Friedmann equation.
However, sum rule $\Omega_\Lambda+\Omega_k=1$ then requires an
empirically improbable large curvature parameter $\Omega_k$. 
Empirical limits are $\Omega_k\simeq\pm0.01$\cite{KOM09}.
\par This issue was examined by solving the modified Friedmann equation
with parameters $k, \beta, \gamma$ set to zero\cite{NES11}. $\Omega_q$ 
is determined by the solution.  The modified sum rule 
$\Omega_\Lambda+\Omega_q=1$ then presents no problem. Computed
$d_L(z)$ agrees with Mannheim's empirical function for $z\leq 1$ to  
graphical accuracy, using parameter $\alpha=\Omega_\Lambda(t_0)=0.732$ 
for $\Omega_k(t_0)=0$. This is consistent with current empirical values
$\Omega_\Lambda=0.726\pm0.015, \Omega_k=-0.005\pm 0.013$\cite{KOM09}.
$\Omega_m$ and $\Omega_r$ can apparently be neglected for $z\leq 1$.
\par $t=0$ is defined by $h(t)=0$ in the conformal model,
which describes an initial inflationary epoch\cite{NES11}.
The modified Friedmann equation was solved numerically for 
$0\leq t\leq t_0$\cite{NES11}, with parameters fitted to $d_L(z)$ 
for $z\leq 1$, to shift parameter $R(z_*)$\cite{WAM07},
and to acoustic scale ratio $\ell_A(z_*)$\cite{WAM07}.
This determines model parameters $\alpha=0.7171, k=-0.01249,
\beta=0.3650\times 10^{-5}$. 
Fixed at $\gamma=3\beta/4R_b(t_0)$, which neglects dark matter,
parameter $\gamma=0.3976\times 10^{-8}$.  There is no significant 
inconsistency with model-independent empirical data\cite{KOM09}.
\par Defining $\zeta=\sxth R-w^2$, the dimensionless sum rule 
determines $\zeta=\xi_0+\xi_1-w^2=h(t)^2(2\Omega_q+\Omega_m+\Omega_r)$. 
For $a\to 0$, when both $\alpha$ and $k$ can be neglected, the 
sum rule implies $\zeta=h(t)^2(\Omega_q+1)$.  For large $a$, 
$\zeta= h(t)^2(2\Omega_q)$.  $\zeta>0$ in both limits, regardless of 
numerical values, since $\Omega_q>0$.  The present empirical parameters
imply that $\zeta$ is positive for all $z$\cite{NES11}.   
\par Conformal symmetry is consistent with any real value of parameter 
$\lambda$.  However, in electroweak theory Higgs symmetry-breaking
requires nonvanishing conformal scalar field $\Phi$\cite{PAS95}.
A positive value of $\zeta$ implies 
\begin{eqnarray}
 \lambda\phi_0^2=\half(w^2-\sxth R)=-\half\zeta<0. 
\end{eqnarray}
As argued above, for $\phi_0^2>0$ this conflicts with existence of the 
hypothetical massive Higgs boson.

\section{Dynamical estimate of parameter $w^2$}
\par Since term $w^2\Phi^\dag\Phi$ in standard parametrized 
$\Delta{\cal L}$ breaks conformal symmetry, it must be generated 
dynamically in a consistent theory\cite{MAN09}.  As shown above, this 
term accounts for dark energy.  Dynamically induced $w^2$ preserves the 
conformal trace condition\cite{NES10}.  
\par The Higgs model deduces gauge boson mass from an exact solution
of the parametrized scalar field equation\cite{CAG98}.  For interacting
fields, this logic can be extended to deduce nominally constant field
parameters from a solution of the coupled field equations.  Such a 
solution of nonlinear equations does not depend on linearization or on
perturbation theory.  
\par Interaction of scalar and gauge boson fields defines a 
quasiparticle scalar field in Landau's sense:
$\Phi$ is dressed via virtual excitation of accompanying gauge fields.
The derivation summarized here considers gravitational field 
$g_{\mu\nu}$ interacting with scalar field $\Phi$ and $U(1)$ gauge 
field $B_\mu$.  Solution of the coupled semiclassical field 
equations\cite{NES10} gives an order-of-magnitude estimate of parameter
$w^2$, in agreement with the empirical cosmological constant, while 
confirming the Higgs formula for gauge boson mass\cite{PAS95,CAG98}.
\par The conformal Higgs model assumes incremental Lagrangian density
$\Delta{\cal L}_\Phi=w^2\Phi^\dag\Phi-\lambda(\Phi^\dag\Phi)^2$,
with undetermined numerical parameters $w^2$ and $\lambda$.
The implied scalar field equation is
$\partial_\mu\partial^\mu\Phi+\sxth R\Phi=
\frac{1}{\sqrt{-g}}\frac{\delta\Delta I}{\delta\Phi^\dag}=
  (w^2-2\lambda\Phi^\dag\Phi)\Phi$.
If $R,w^2,\lambda$ are constant, this has an exact solution $\Phi^\dag\Phi=\phi_0^2=(w^2-\sxth R)/2\lambda$,
if this ratio is positive.
For massive complex vector field $B_\mu$, parametrized 
$\Delta{\cal L}_B$ implies field equation
$\partial_\nu B^{\mu\nu}=
2\frac{1}{\sqrt{-g}}\frac{\delta\Delta I}{\delta B_\mu^*}=
m_B^2 B^\mu-J_B^\mu$. 
\par For interacting fields, both $\Delta{\cal L}_\Phi$ and
$\Delta{\cal L}_B$ can be identified with incremental
Lagrangian density $\Delta{\cal L}=$ 
\begin{eqnarray}
 \frac{i}{2}g_b B^\mu(\partial_\mu\Phi)^\dag\Phi
-\frac{i}{2}g_b B_\mu^\dag\Phi^\dag\partial^\mu\Phi
+\frth g_b^2\Phi^\dag B_\mu^\dag B^\mu\Phi,
\end{eqnarray}
due to covariant derivatives, with coupling constant $g_b$.
Evaluated for solutions of the coupled field equations,
\begin{eqnarray}
 2\frac{1}{\sqrt{-g}}\frac{\delta\Delta I}{\delta B_\mu^*}=
\half g_b^2\Phi^\dag\Phi B^\mu-ig_b\Phi^\dag\partial^\mu\Phi 
\end{eqnarray}
implies Higgs mass formula $m_B^2=\half g_b^2 \phi_0^2$.
The fields are coupled by current density
$J^\mu_B=ig_b\Phi^\dag\partial^\mu\Phi$.
For the scalar field, neglecting derivatives of $B_\mu$,
\begin{eqnarray}
 \frac{1}{\sqrt{-g}}\frac{\delta\Delta I}{\delta \Phi^\dag}=
\frth g_b^2B_\mu^*B^\mu \Phi
-\frac{i}{2}g_b(B_\mu^*+B_\mu)\partial^\mu\Phi
\end{eqnarray}
implies $w^2=\frth g_b^2 B_\mu^*B^\mu$.
\par For $\zeta=\sxth R-w^2>0$, 
$\Phi^\dag\Phi=\phi_0^2=-\zeta/2\lambda$
solves the scalar field equation if $\lambda<0$.
Ricci scalar $R(t)$ varies on a cosmological time scale, so that
$\frac{{\dot\phi}_0}{\phi_0}=\half\frac{{\dot R}}{R-6w^2}\neq0$, for
constant $w^2$ and $\lambda$.  This implies small but nonvanishing
real $\frac{{\dot\phi}_0}{\phi_0}$, 
hence nonzero pure imaginary source current density 
$J^0_B=ig_b\phi_0^*\partial^0\phi_0
 =ig_b\frac{{\dot\phi}_0}{\phi_0}\phi_0^2$.
\par Derivatives due to cosmological
time dependence act as a weak perturbation of SU(2) scalar field 
solution $\Phi=(\Phi_+,\Phi_0)\to (0,\phi_0)$.
Neglecting extremely small derivatives of the induced gauge fields
(but not of $\Phi$), the gauge field equation reduces to
$m_B^2 B^\mu= J_B^\mu$. 
Implied pure imaginary $B^\mu$ does not affect parameter $\lambda$.
The coupled field equations imply $w_B^2=\frth g_b^2|B|^2$,
proportional to $(\frac{{\dot\phi}_0}{\phi_0})^2$.
Since observable properties depend only on $|B|^2$, a pure imaginary 
virtual field implies no obvious physical inconsistency.
Gauge symmetry is broken in any case by a fixed field solution.
The scalar field is dressed by the induced gauge field.
\par Numerical solution of the modified Friedmann 
equation\cite{NES11,NES10} implies 
$\zeta(t_0)=1.224\times 10^{-66}eV^2$, at present time $t_0$.
Given $\phi_0=180GeV$\cite{CAG98}, 
$\lambda=-\half\zeta/\phi_0^2=-0.189\times 10^{-88}$.
\par U(1) gauge field $B_\mu$ does not affect $\lambda$. 
Using $|B|^2=|J_B|^2/m_B^4,
|J_B|^2=g_b^2(\frac{{\dot\phi}_0}{\phi_0})^2\phi_0^4$ and
$m_B^2=\half g_b^2\phi_0^2$, 
the dynamical value of $w^2$ due to $B_\mu$
is $w_B^2=\frth g_b^2|B|^2=(\frac{{\dot\phi}_0}{\phi_0})^2$. 
\par From the solution of the modified Friedmann equation\cite{NES10},
$\frac{{\dot\phi}_0}{\phi_0}(t_0)=-2.651$ and $w_B^2=7.027$,
in Hubble units, so that
$w_B=2.651\hbar H_0=3.984\times 10^{-33}eV$ in energy units.
This can be considered only an order-of-magnitude estimate, since
time dependence of the assumed constants, implied by the present theory,
was not considered in fitting empirical cosmological data\cite{NES11}.  
Moreover, the SU(2) gauge field has been omitted.

\section{Note on dark matter}
As stated in\cite{NES11}, interpretation of parameter $\Omega_m$
may require substantial revision of the standard cosmological model.
Directly observed inadequacy of Newton-Einstein gravitation may imply
the need for a modified theory rather than for inherently
unobservable dark matter.
\par Mannheim has applied
conformal gravity to anomalous galactic rotation\cite{MAN06},
fitting observed data for a set of galaxies covering a large range
of structure and luminosity.  The role played in standard
$\Lambda$CDM by dark matter, separately parametrized for each galaxy,
is taken over in conformal theory for Schwarzschild geometry
by an external linear radial potential.  The remarkable fit to observed
data shown in\cite{MAN06}[Sect.6.1,Fig.1] requires only two universal
parameters for the whole set of galaxies. 
\par As discussed by Mannheim\cite{MAN06}[Sects.6.3,9.3],
a significant conformal contribution to centripetal acceleration is
independent of total galactic luminous mass.  This implies an external
cosmological source.  Such an isotropic source would determine 
an inherently spherical halo of gravitational field surrounding
any galaxy.  Quantitative results for lensing and for galactic
clusters should be worked out before assuming dark matter.

\section{Conclusions}
This paper is concerned with determining parameters $w^2$ and $\lambda$
in the incremental Lagrangian density of the Higgs model, 
$\Delta{\cal L}_\Phi=(w^2-\lambda\Phi^\dag\Phi)\Phi^\dag\Phi$.
Fitting the modified Friedmann equation to cosmological 
data\cite{NES11} implies dark energy parameter 
$\Omega_\Lambda=w^2=0.717$, so that empirical
$w=\sqrt{0.717}\hbar H_0=1.273\times 10^{-33}eV$.
\par The modified Friedmann equation determines the time derivative of
the cosmological Ricci scalar, which implies nonvanishing source current
density for induced U(1) gauge field $B_\mu$, treated here 
as a classical field in semiclassical coupled field equations. 
The resulting gauge field intensity estimates the U(1) contribution
to $w^2$ such that $w_B=2.651\hbar H_0=3.984\times 10^{-33}eV$.
This order-of magnitude agreement between computed $w_B$ and empirical
$w$ supports the conclusion that conformal theory explains both the 
existence and magnitude of dark energy\cite{NES10}.
\par The present argument obtains an accurate empirical value of
parameter $\lambda$ from the known dark energy parameter\cite{KOM09},
from the implied current value of Ricci scalar $R$\cite{NES11},
and from scalar field amplitude $\phi_0$ determined by gauge
boson masses\cite{CAG98}.
The mass parameter for a fluctuation of the
conformal Higgs scalar field satisfies $m_H^2=4\lambda\phi_0^2$.
Empirical value $\lambda=-0.189\times 10^{-88}$ is negative,
implying finite pure imaginary parameter $m_H$.  If such a particle or 
field could exist or be detected, this would define a 
tachyon\cite{FEI67}, the quantum version of a classical particle that 
moves more rapidly than light.  Experimental data rule out a standard 
massive Higgs boson with mass $0\leq m_H\leq108$GeV\cite{DGH89,OPA10}.
However, a Higgs tachyon\cite{FEI67} might either not exist at all, 
or elude detection in experiments designed for a classical massive 
Higgs boson.  The present results would only be inconsistent if 
experimental Higgs searches to date were capable of detecting a Higgs 
tachyon and failed to do so. Conformal theory clearly rules out a 
standard Higgs boson in the multi-GeV range.
%**************** VM screen width *************************************%

\end{document}